\def   \ni {\noindent}

\def   \sssk {\vskip  3truept}
\def   \ssk {\vskip  5truept}

\def   \bsk {\vskip 15truept}

\def   \newline {\hfil\break}

\documentstyle[psfig]{article}
\begin{document}

\hsize 5truein
\vsize 8truein
\font\abstract=cmr8
\font\keywords=cmr8
\font\caption=cmr8
\font\references=cmr8
\font\text=cmr10
\font\affiliation=cmssi10
\font\author=cmss10
\font\mc=cmss8
\font\title=cmssbx10 scaled\magstep2
\font\alcit=cmti7 scaled\magstephalf
\font\alcin=cmr6 
\font\ita=cmti8
\font\mma=cmr8
\def\ref{\par\noindent\hangindent 15pt}
\null
\baselineskip = 12pt
\def\ol{\Omega_{\Lambda}}
\def\ob{\Omega_{b}}
\def\obh{\Omega_{b}h^{2}}
\def\oc{\Omega_{CDM}}
\def\om{\Omega_{m}}
\def\on{\Omega_{\nu}}
\def\og{\Omega_{\gamma}}
\def\os{\Omega_{stars}}
\def\od{\Omega_{dust}}
\def\ogas{\Omega_{gas}}
\def\t{t_{o}}
\def\gtrsim{\mathrel{\hbox{\rlap{\hbox{\lower4pt\hbox{$\sim$}}}\hbox{$>$}}}}
\def\lesssim{\mathrel{\hbox{\rlap{\hbox{\lower4pt\hbox{$\sim$}}}\hbox{$<$}}}}
\def\lsim   {\wisk{_<\atop^{\sim}}}
\def\gsim   {\wisk{_>\atop^{\sim}}}
\def\etal{{\em et al.~}}
\def\apj{{\em Ap.J.}}
\def\apjl{{\em Ap.J.L.}}
\def\mnras{{\em M.N.R.A.S.}}
\def\aa{{\em A \& A}}
\def\be{\begin{equation}}
\def\ee{\end{equation}}


\title{\ni \begin{center} What is the Universe made of?\\
How old is it?\\

\end{center}}  

\bsk
\author{\ni Charles H. Lineweaver}\\            
\sssk
\affiliation{University of New South Wales} 
\baselineskip = 12pt
\ssk

\abstract{ABSTRACT\\
{\it For the past 15 years most astronomers have assumed that 95\% of the Universe was
in some mysterious form of cold dark matter. They also assumed that the cosmological 
constant, $\ol$, was Einstein's biggest blunder and could be ignored.
However, recent measurements of the cosmic microwave background combined with other
cosmological observations strongly suggest that $75\%$ of the Universe is made of
cosmological constant (vacuum energy), while only $20\%$ is made of non-baryonic cold 
dark matter. Normal baryonic matter, the stuff most physicists study, makes up about
$5\%$ of the Universe. If these results are correct, an unknown 75\% of the Universe has 
been identified. Estimates of the age of the Universe depend upon what it is made of. 
Thus, our new inventory gives us a new age for the Universe: $13.4 \pm 1.6$ Gyr.}}

\baselineskip = 12pt

\begin{quote}
\indent ``The history of cosmology shows us that in every age devout people believe\\
 that they have at last discovered the true nature of the Universe.''\\
(E. Harrison in  {\bf Cosmology: The Science of the Universe} 1981)
\end{quote}

\section{Progress}

A few decades ago cosmology was laughed at for being
the only science with no data. Cosmology was theory-rich but data-poor.
It attracted armchair enthusiasts spouting 
speculations without data to test them. It was the only science 
where the errors could be kept in the exponents -- where you could set the speed of
light $c=1$, not for dimensionless convenience, but because the observations 
were so poor that it didn't matter. The night sky was calculated to be as
bright as the Sun and the Universe was younger than the Galaxy.

Times have changed. We have entered a new era of precision cosmology.
Cosmologists are being flooded
with high quality measurements from an army of new instruments
\footnote{COBE, ISO, IRAS, HIPPARCOS, HST, IUE, BeppoSax, UHURU, ROSAT,  Chandra, BATSE, 
VLA, ATCA, Arecibo, KAO, SOFIA, SCUBA, BIMA, KECK, VLT, CFHT,  MMT, 
UKIRT, AAT,  CTIO, FLY's EYE,
CSO, JCNT, NTT, KPNO, UKIRT, INT, JKT, WHT,
Magellan, GTC,  LBT, MAX,
 Kamiokande, Super Kamiokande, HOMESTAKE,
VIRGO, LIGO,
Gravity Probe-B, GINGA, ASTRO A,B,C,D, CERN, FERMILAB, STANFORD, DS1, MILAGRO, Gran Sasso, SNO...}
.
We are observing the Universe at new frequencies, with higher sensitivity, higher 
spectral resolution and higher spatial resolution.  We have so much new data that 
state-of-the-art computers process and store them with difficulty.
Cosmology papers now include error bars -- often asymmetric and
sometimes even with a distinction made between statistical and systematic error bars. 
This is progress.

\clearpage

The standard hot big bang model describes the evolution of the Universe. 
It is the dominant paradigm against which all new ideas are tested.
It provides a consistent framework into which all the relevant cosmological data seem to fit.
Progress has been made in working out the details of this hot big bang model -- for
it is the details which provide new, unprecedentedly precise answers to questions of mythical importance:
What is the Universe made of? How old is the Universe?

\begin{figure}[!h]
\centerline{\psfig{figure=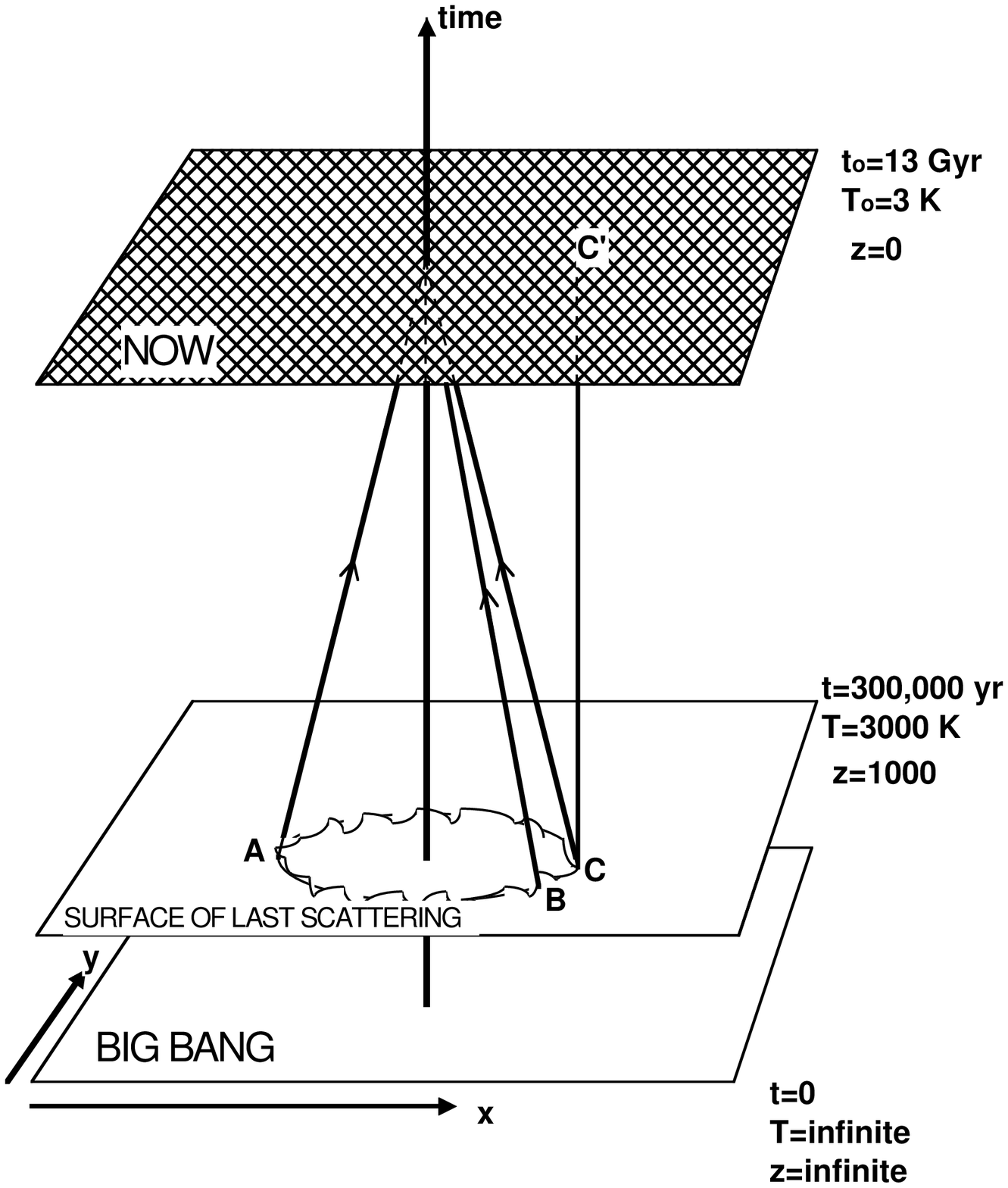,angle=0,height=11.0cm,width=12cm,rheight=11.0cm,rwidth=12cm}}
\caption{{\bf Figure 1. Our view of the Universe}\\
In this spacetime diagram we are at the apex of our past light cone.
All the photons we see come to us along the surface of this cone.
One spatial dimension has been suppressed.
When we look as far away as we can, we see the oldest observable photons
-- the CMB -- coming from the wavy circle in the surface of last scattering
(A, B and C are on the circle).
The opaque surface of last scattering is the boundary between the current
transparent universe and the hotter, denser, opaque, ionized  universe.
The figure gives the time, temperature and
redshift of the big bang, the surface of last scattering and today.
This is a comoving diagram, that is, the expansion of the Universe is not shown.
We see the object C on the surface of last scattering as it was 13 Gyr ago.
Today C has become C', but since the speed of light is not infinite, we cannot see
C' now.
} \label{fig-1}
\end{figure}

\clearpage
\section{The CMB: cosmology's coolest new tool.}

The cosmic microwave background (CMB) is the oldest fossil we have ever 
found. It is a bath of photons coming from every direction. 
These photons are the afterglow of the big bang.
Their long journey toward us has lasted more than 99.99\% of the 
age of the Universe and began when the Universe
was one thousand times smaller than it is today.
The CMB was emitted by the hot plasma of the Universe long before 
there were planets, stars or galaxies.
The CMB is an isotropic field of electromagnetic
radiation -- the redshifted relic of the  hot big bang.

One of the most recent and most important advances in astronomy has been the discovery 
of hot and cold spots in the CMB based on data from the COBE satellite (Smoot \etal 1992).
This discovery has been hailed as ``Proof of the Big Bang''
and the ``Holy Grail of Cosmology'' and elicited comments like:
``If you're religious it's like looking at the face of God''
(George Smoot) and
``It's the greatest discovery of the century, if not of all time''
(Stephen Hawking).
As a graduate student analyzing COBE data at the time,
I knew we had discovered something fundamental but
its full import didn't sink in until one night after
a telephone interview for BBC radio. I asked the interviewer for
a copy of the interview, and he told me that would be
possible if I sent a request to the {\it religious affairs} department.

The CMB comes from the surface of last scattering of the Universe.
When you look into a fog, you are looking at a surface of
last scattering. It is a surface defined by all the molecules
of water which scattered a photon into your eye.
On a foggy day you can see 100 meters, on really foggy days
you can see 10 meters. If the fog is so dense that you cannot see 
your hand then the surface of last scattering 
is less than an arm's length away.
Similarly, when you look at the surface of the Sun you are seeing
photons last scattered by the hot plasma of the photosphere.
The early Universe is as hot as the Sun and similarly
the early Universe has a photosphere (the surface of last scattering) 
beyond which (in time and space) we cannot see (Fig. 1). 
As its name implies, the surface of last scattering is where
the CMB photons were scattered for the last time before
arriving in our detectors.
The `surface of last screaming' presented in Fig. 2 is a pedagogical
analog.


Since the COBE discovery of hot and cold spots in the CMB, 
anisotropy detections have been reported 
by more than a dozen groups with various instruments, at various frequencies and 
in various patches and swathes of the microwave sky. 
Fig. 3 is a compilation 
of recent measurements. The COBE measurements (on the left) are at large 
angular scales while most recent measurements are trying to constrain the 
angular scale and amplitude of the dominant first peak at $\sim 1/2$ degree (the size of the full Moon).
This dominant peak and the smaller amplitude peaks at smaller angular scales are due
to acoustic oscillations in the photon-baryon fluid in cold dark matter (CDM) gravitational potential wells.
The detailed features of these peaks in the power spectrum are dependent on a large number  of
cosmological parameters including,

\begin{description}
\item {$\mathbf \om$} the density of matter (where $\om = \oc + \ob$)
  \begin{itemize}
     \item {$\mathbf \oc$} the density of cold dark matter
     \item {$\mathbf \ob$} the density of normal baryonic matter
  \end{itemize}
\item {$\mathbf \ol$} the density of vacuum energy (cosmological constant)
\item{$\mathbf h$} the Hubble constant (giving the rate of expansion of the Universe)
\end{description}

\begin{figure}
\centerline{\psfig{figure=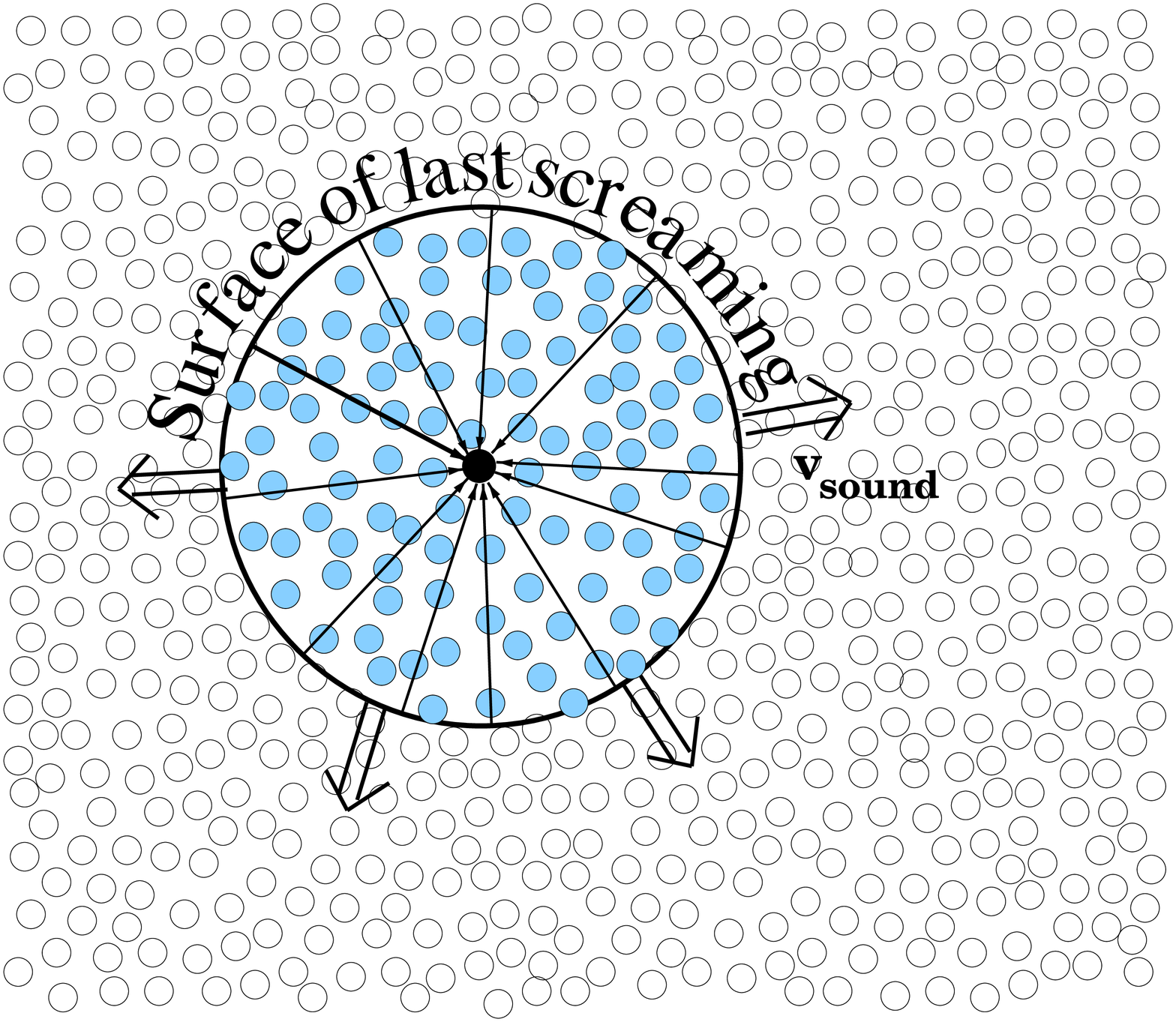,angle=0,height=10cm,width=11cm,rheight=11cm,rwidth=11cm}}
\caption{{\bf Figure 2. The Surface of Last Screaming.}\\
Consider an infinite field full of people screaming. 
The circles are their heads. 
You are screaming too. (Your head is the black dot.)
Now suppose everyone stops screaming at the same time. 
What will you hear? Sound travels at 330 m/s.
One second after everyone stops screaming you will be able to hear the screams from a `surface of last screaming' 330 meters away from you in all directions.
After 3 seconds the faint screaming will
be coming from 1 km  away...etc. 
No matter how long you wait, faint screaming will always be coming from 
the surface of last screaming -- a surface that is receeding from you
at the speed of sound (`v$_{sound}$').
The same can be said of any observer -- each is the center of a surface
of last screaming.
In particular, observers on your surface of last screaming are currently hearing you scream
since you are on their surface of last scattering.
The screams from the people closer to you than the surface of last
screaming  have passed you by -- you hear nothing from them (grey heads).
When we observe the CMB in every direction
we are seeing photons from the surface of last scattering. 
We are seeing back to a time soon after the big bang when the entire
Universe was opaque (screaming).
If the Universe were not expanding, the surface of last scattering would be
receding at the speed of light. The expansion of the Universe adds
an additional recession velocity and makes the surface of last scattering
recede at $\sim 3 \: c$. 
} \label{fig-2}
\end{figure}
%
\clearpage
\begin{figure}
\centerline{\psfig{figure=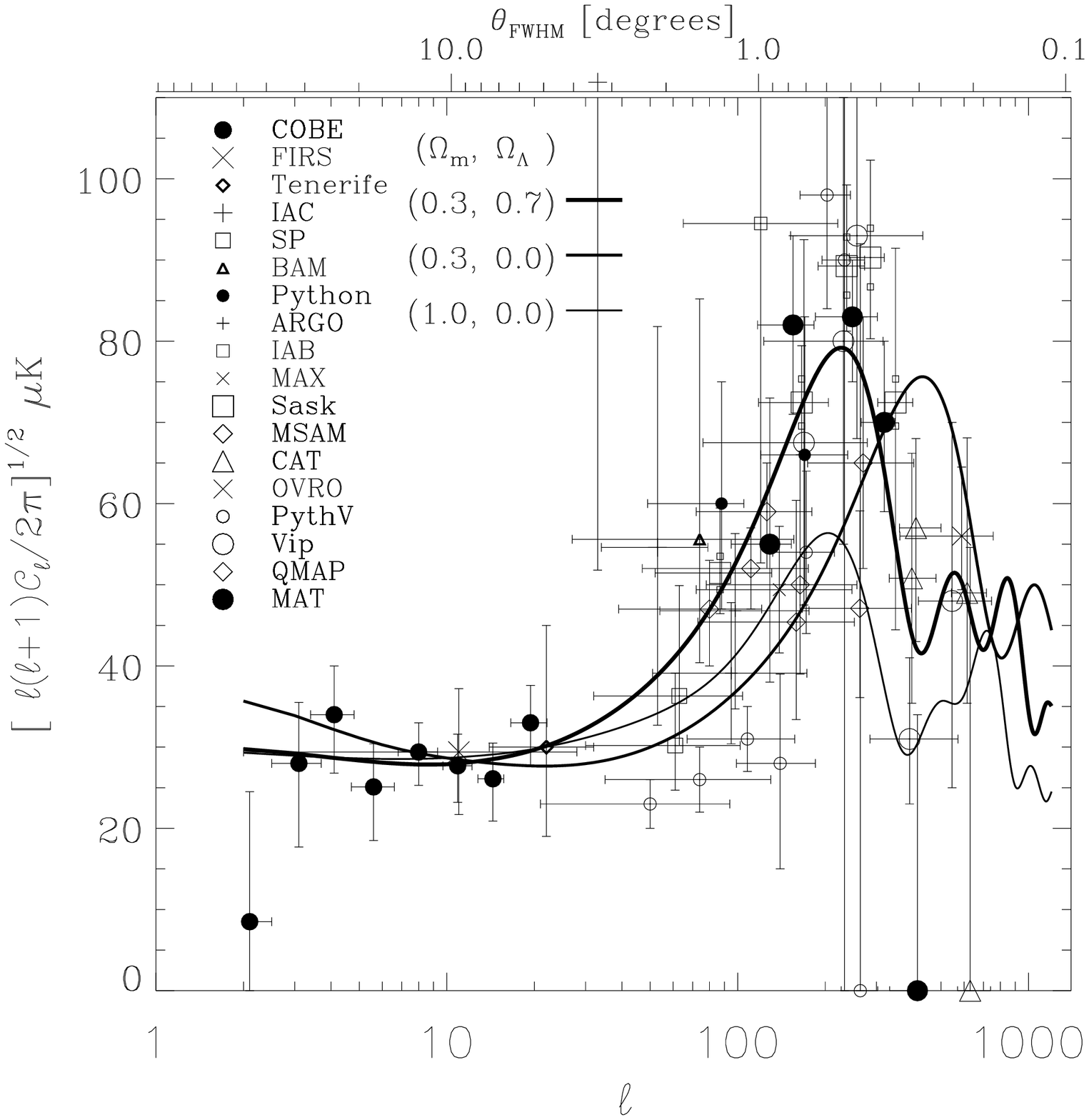,angle=0,height=12cm,width=12cm}}
\caption{{\bf Figure 3.  Measurements of the CMB power spectrum.}\\
The amplitudes of the hot and cold spots in the CMB depend on their angular
size. Angular size is noted in degrees on the top x axis.
The y axis is the rms amplitude of the temperature fluctuations.
For example, the fluctuations at  1/2  degree are at 90 $\mu$K while
the fluctuations at 10 degrees are at 30 $\mu$K.
This means that at the 1/2 degree angular scale, the hot spots are 3 times hotter and the cool spots are 
3 times cooler than at the 10 degree scale.
Each CMB experiment is sensitive only to a limited range of angular scale.
When the measurements at various angular scales are put together they form the CMB power spectrum.
The COBE-DMR points at $\theta_{FWHM} \gtrsim 7^{o}$ are primordial fluctuations
corresponding to scales so big they are `non-causal', i.e., they have physical sizes 
larger than the distance light could have travelled between the big bang and their 
age at the time we see them (300,000 years after the big bang).
They are either the initial conditions of the Universe or were laid down
during an epoch of inflation $\sim 10^{-35}$ seconds after the big bang. 
New sets of points are being added  every month or so. 
The three curves represent the three most popular models.
$\chi^{2}$ fits of this data to such model curves yields the constraints in Fig. 5A.}
%
\label{fig-3}
\end{figure}
\clearpage

My work over the past few years has been to extract values for these
parameters by comparing the most recent measurements of the CMB with
parameter-dependent models (Lineweaver \& Barbosa 1998, Lineweaver 1998, 1999a).
The three curves in Fig. 3 are examples of such models and represent the three most
popular candidates for the best fit to reality. They are known as
standard-CDM, Open-CDM and  $\Lambda$CDM:
$(\Omega_{m},\Omega_{\Lambda}) = (1.0,0.0),(0.3,0.0)$ and $(0.3, 0.7)$ respectively.
The principal support for these models comes from theory, tradition and data, respectively.
The $\Lambda$CDM model fits the position and amplitude of the dominant
first peak quite well. 
The standard-CDM model has a peak amplitude much too low. 
The open-CDM model has the peak at angular scales too small to fit the data 
and is strongly excluded by a fuller analysis (see Fig. 5A).

%

\section{What is the Universe made of?}

If we know what the Universe is made of, we know how it will behave and how it has behaved -- we know its
dynamics and shape and destiny -- whether it will expand forever or collapse in a big
crunch -- whether it is spatially finite or infinite -- whether it is  10 billion years old or 20.
Many of these issues can be reduced to the question: Where does our Universe lie
in the ($\om, \ol$) plane?  Observational constraints in this plane are then  the
crucial arbiters. Figure 4 can be used to translate $\om$ and $\ol$ constraints into the words
most commonly used to describe the Universe.

In cosmology we keep track of the components of the Universe by their
densities: $\om$, $\oc$, $\ob$, $\ol$. These are all dimensionless densities 
expressed in units of the critical density,
$10^{-29}$ g cm$^{-3}$ (9 orders of magnitude emptier than the best laboratory
vacuums). 
If the Universe has the critical density ($\om = 1$), then its current rate of expansion
is analogous to the escape velocity, that is, it will expand forever, 
asymptotically approaching no expansion
as $t \rightarrow \infty$ (just as the velocity of an object with escape
velocity asymptotically approaches 0). 
One can read from Fig. 4 that an $(\om,\ol) = (1.0,0.0)$ universe is flat, 
decelerating and will expand forever.


\subsection{Much Ado About Nothing}


One of the most surprising recent advances in cosmology is that 75\% of the Universe
seems to be made out of nothing, i.e., the energy of the vacuum. 
I have assembled much of the observational evidence for this in Fig. 5.
Recent CMB anisotropy measurements favour the elongated triangle in panel A
of Fig. 5. This plot shows that if $\ol = 0$ then $\om \sim 0.3$ is 
more than $\sim 4 \sigma$ from the best fit and $\om \sim 0.1$ is more than $\sim 7 \sigma$
away. The confidence levels in this diagram are very rough but the message is clear:
if $\ol = 0$, then low $\om$ models are strongly excluded by the CMB data.
No other data set can exclude this region with such high confidence.
The combination of CMB and supernovae constraints (Fig.5B) provides strong evidence 
that $\ol > 0$. If any $\ol = 0$ model can squeak by the new supernovae
constraints it is the very low $\om$ models.
However these models are the ones most strongly excluded by the CMB data.
The constraints shown in panels C, D and E support this result.
Separately these data sets 
cannot determine unambiguously what the destiny of the Universe will be. However, together 
they form a powerful interlocking network of constraints yielding the most precise 
estimates of $\om$ and $\ol$. The result is strong evidence and the best
evidence to date that the Universe will expand forever, dominated by a 75\% contribution
from the vacuum.

I believe this result is robust because of a series of
conservative choices made in the analysis and because
it arises when the data sets are combined individually 
(as in panels B,C,D and E) or combined together (as in F).
Systematic errors may compromise one or the other of the 
observations but are less likely to bias all of the 
observations in the same way.

The $\Lambda$CDM region of the ($\om, \ol$) plane  fits the CMB, supernovae and other 
data sets and should be viewed as the new standard model of cosmology.
Standard CDM with $\om = 1$ and $\ol = 0$ is a simpler model, but circular
planetary orbits are also simpler than ellipses.
%
The results presented in Fig. 5F (Lineweaver 1999a) quantify the main components
of the new standard $\Lambda$CDM model.
They are  depicted in Fig. 6 and are as follows:

\begin{center}
{\bf Table of Contents of the Universe}
\end{center}

\begin{itemize}
\item  {\bf 75\% Vacuum energy, cosmological constant, $\ol$}
 
The vacuum of modern physics is not empty. It is seething with virtual particles coming in and out of 
existence. All this seething produces a vacuum energy  (the zero point energy of quantum field theory)
which has a negative pressure. 
Unlike normal mass which slows down the expansion of the Universe, vacuum energy speeds 
up the expansion. It's a bit like discovering compressed springs everywhere in the vacuum 
of space. These springs make the Universe expand.
This mysterious stuff does not clump. It is the Lorentz invariant structure of the vacuum and its existence
is probably most directly established by the Casimir effect and the Lamb shift.
$\ol = 0.65 \pm 0.13$ corresponding to $74 \pm 4 \%$ of the Universe.

\item {\bf 20\% Cold Dark Matter (CDM)}\\
Non-baryonic and non-relativistic, CDM density fluctuations
collapse gravitationally. It clumps. Corresponding CDM potential wells (and hills) 
produce the hot and cold spots in the CMB and are the principle seeds for the 
formation of the large scale structure we see around us today (galaxies, great walls, voids etc).
This non-baryonic stuff has never been detected directly.
$\oc = 0.19 \pm 0.09$ corresponding to $21 \pm 7 \%$ of the Universe.
Leading candidates for it are axions or neutralinos (see Turner 1999).

\item {\bf 5\% Normal baryonic matter}\\
This is the normal stuff that stars and ourselves are made of. We breathe it, eat it and physicists
study it.  $\ob = 0.04 \pm 0.02$, corresponding to $5\pm 2 \%$ of the Universe.
This value comes from big bang nucleosynthesis calculations and deuterium measurements
in quasar spectra.  In terms of elemental composition this normal baryonic matter is
75\% hydrogen, 
23\% helium,  
and 2\% all other elements.  
In terms of phase (see Fukugita \etal 1998, Cen \& Ostriker 1998), it is
80\% diffuse hot ionized gas,  
17\% stars                   
and 3\% neutral gas and dust.     

\end{itemize}

The total density of the Universe is $\Omega_{total} = \om + \ol = 0.88^{+0.07}_{-0.10}$ (Fig. 5F). 
The percentages listed above are based on $\Omega_{total} = 0.88$ (not $\Omega_{total} = 1$,  most versions of inflation have $\Omega_{total} = 1$).
The density from photons (from the CMB and from stars) is negligible: $0.006\%$ of the Universe.
I have left out one ingredient of the universe because we don't know whether it is important or not -- neutrinos.
We know their number density fairly accurately. It's 
\clearpage

\begin{figure}
\centerline{\psfig{figure=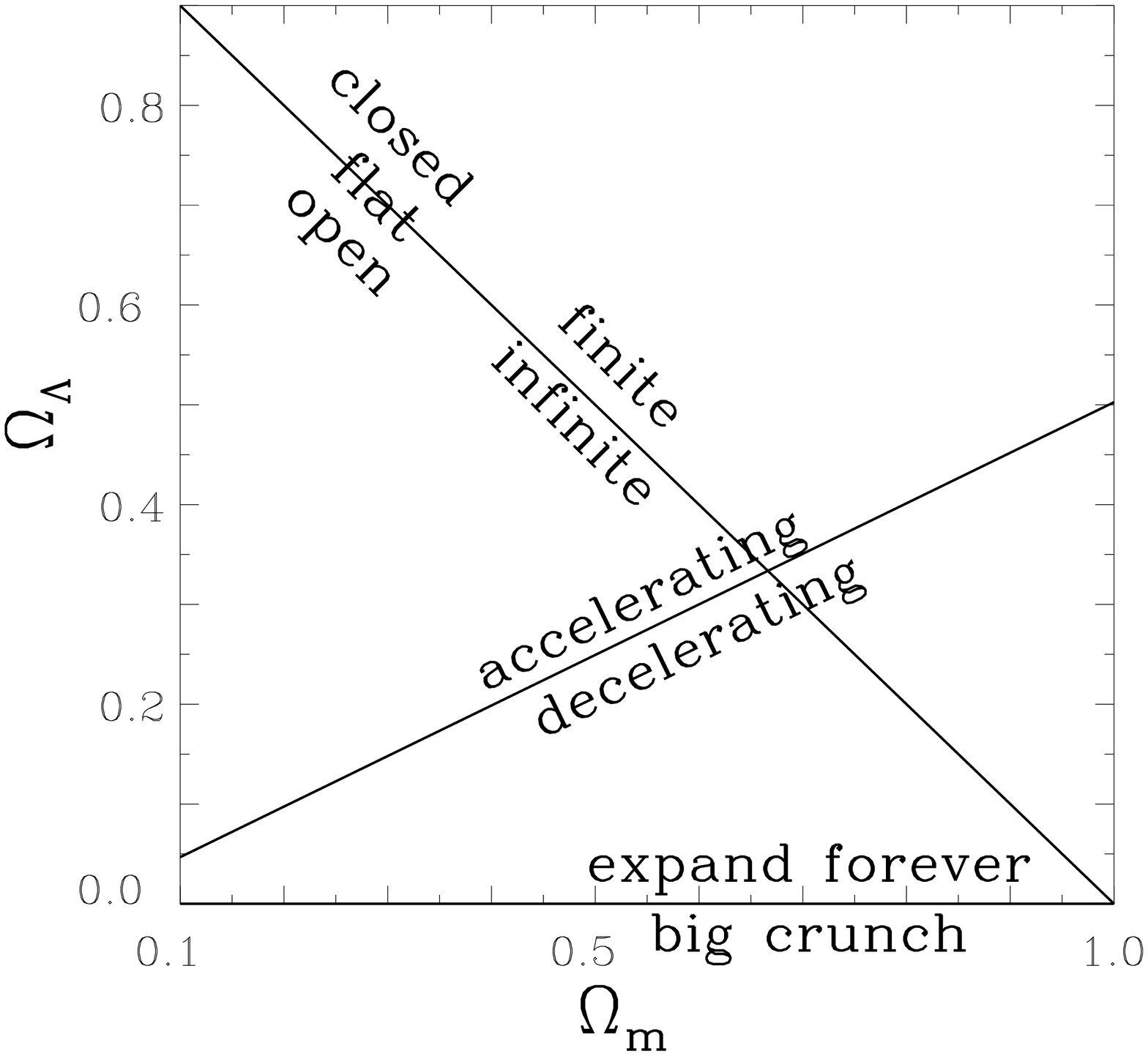,angle=0,height=14cm,width=13cm,rheight=13cm}}
\caption{{\bf Figure 4. Describing the Universe.}\\
The language used to describe the Universe, e.g.
`infinite/finite', `open/flat/closed', `accelerating/decelerating' and `a universe which will
expand forever/end in a big crunch', can be confusing. 
However the boundaries between these various types of universe can be
simply represented in the ($\om, \ol$) plane.
For example, spatially open universes (3-D analog of the surface of a saddle, 
negative curvature) are in the lower left
while spatially closed universes (3-D analog of the surface of a sphere, 
positive curvature) are in the upper right.
Flat Euclidean universes are on the diagonal line between them.
Flat and open models are spatially infinite; closed models are finite.
Notice that one can have finite universes which expand forever and can be either accelerating or
decelerating. One can also have infinite universes which collapse into a big crunch 
(if $\ol < 0$).
A detail that is slightly ambiguous: if $\ol = 0$ then $\om \le 1$ universes expand forever
while $\om > 1$ universes crunch.
Observational constraints in this ($\om, \ol$) plane are given in Fig. 5;
they favour accelerating, slightly open, but nearly flat universes with $\om \approx 0.3$
and $\ol \approx 0.7$.
} \label{fig-4}
\end{figure}

\begin{figure}
\centerline{\psfig{figure=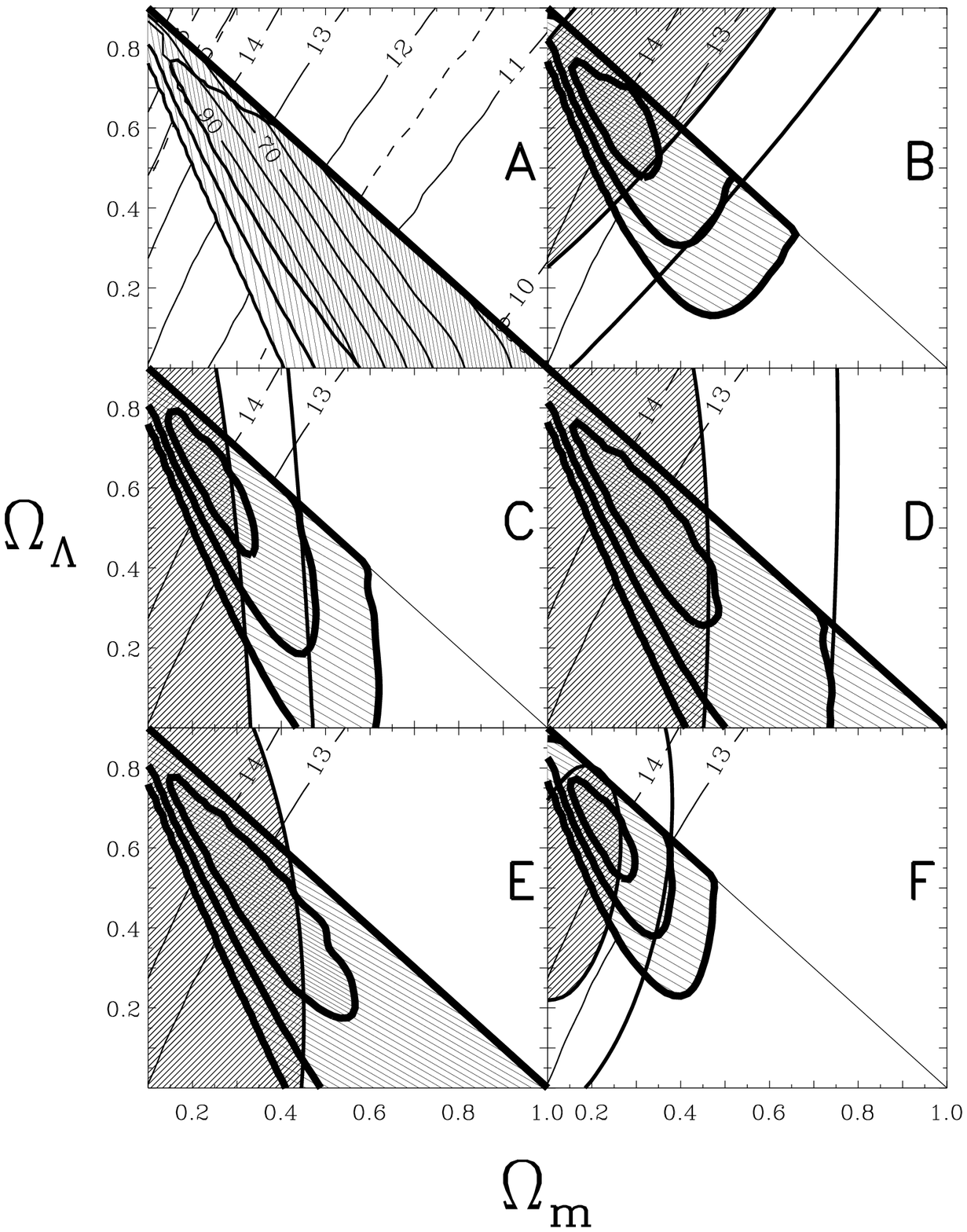,angle=0,height=13.5cm,width=14cm,rheight=14cm,rwidth=14cm}}
\caption{{\bf Figure 5. Observational constraints in the ($\om, \ol$) plane.}\\
These 6 panels show the regions of the ($\ol,  \om$) plane preferred by the data.
The CMB constraint is in the top left panel (A).
Other constraints are from type Ia supernovae (B),
galaxy cluster mass-to-light ratios (C),
galaxy cluster evolution (D) and double lobed radio 
sources (E) and all combined (F).
The thickest contours in each panel are from combining each constraint
with the CMB constraint from A.
The combined constraints (F) yield
$\ol = 0.65 \pm 0.13$ and $\om = 0.23 \pm 0.08$ and
thus $\Omega_{total} = 0.88^{+0.07}_{-0.10}$.
$\Lambda$CDM models in the upper left
are consistent with all the data sets.
The CMB excludes the lower left region of the ($\om, \ol$) plane while each of the 
other constraints excludes the lower right. 
In A, the contours labeled `10' through `14' (Gyr) are the iso-age contours for a Hubble constant $h = 0.68$; 
the 13 and 14 Gyr contours are repeated in all panels.
The contours within the CMB 68\% CL are the best-fitting $H$ values.
See Lineweaver (1998,1999a) for details. 
} \label{fig-5}
\end{figure}
%
\begin{figure}[!h]
\centerline{\psfig{figure=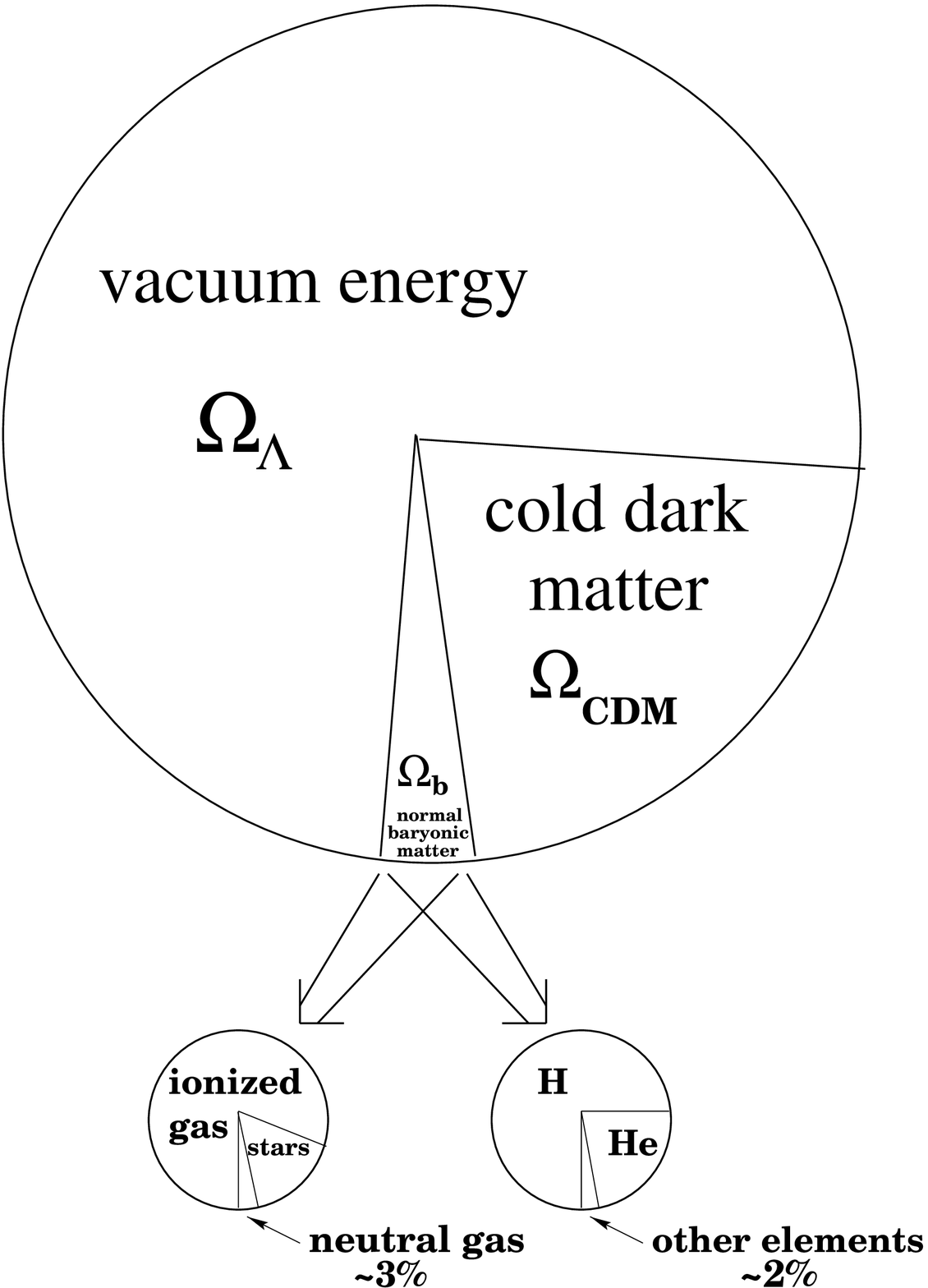,angle=0,height=15cm,width=11cm,rheight=15.3cm,rwidth=11cm}}
\caption{{\bf Figure 6. What is the Universe made of?}\\
$\ol (75\%)$ is now controlling the dynamics of the Universe, causing the Universe to accelerate.
$\oc (20\%)$ acts in the opposite sense, trying to decelerate the Universe (slow 
down the rate of expansion) but it can't compete with $\ol$ (see Fig. 8). $\ob (5\%)$ is a  pawn,  
pushed and pulled around by the gravitational potentials due to spatial variations in the density of CDM.
Most physicists study the 5\% of the Universe made up of normal baryonic matter.
Taking a closer look at the baryonic matter, 98\% of it is either hydrogen or helium and
80\% of it is in difficult to detect ionized gas.
See the Table of Contents of the Universe (p. 7) for details.
} \label{fig-6}
\end{figure}

\clearpage
\noindent
the uncertainty in their mass which is responsible for our ignorance. 
Much effort is being put into measuring the mass(es) of neutrinos.
Potentially they could contribute  more
than all the baryons and probably as much as all the stars.
A good guess might be $\on = 0.05_{-0.047}^{+0.10}$ where the upper limit comes from
the tendency of relativisitc particles to escape from small scale structures, i.e.,
if neutrinos formed more than 15\% of the Universe, we could see much less
small scale structure. The lower limit comes from the recent Super Kamiokande
detection of a small but positive mass difference between two neutrino species
(Fukuda \etal 1998). 
They could be negligible at $0.3\%$ of the Universe or they could be $\sim 15\%$ of the Universe (Turner 1999).
A $\sim 10\%$ contribution would make $\Omega_{total} \sim 1$ as preferred by inflation and would reduce
the contribution of $\ol$ from 75\% to 65\%.

The values quoted above for the composition of the Universe are not 
universally accepted.
A vocal minority of $\Lambda$-phobic cosmologists and particle
theorists believe that any $\ol > 0$ result has got to be wrong.
Their reasoning goes something like this. Theory predicts that $\ol \gtrsim 10^{52}$. Since it is 
obviously not this value, $\ol$ must be zero based on supersymmetric cancellation of the contributions
to the vacuum energy from bosons and fermions.
See Cohn (1998, Section II) for a more judicious discussion.

\section{How old is the Universe?}

In the big bang model, the age of the Universe, $\t$, is a function of three
parameters: $h$, $\om$ and $\ol$. The dimensionless Hubble's constant, $h$,
tells us how fast the Universe is expanding. The  matter density $\om$ slows the expansion while 
the vacuum energy $\ol$ speeds up the expansion.
Until recently, large uncertainties in the measurements of $h$, $\om$ and $\ol$
made efforts to determine $\t(h,\om,\ol)$ unreliable.
Theoretical preferences were, and still are, often used to remedy these
observational uncertainties. One assumed the standard model
($\om = 1$, $\ol = 0$), dating the age of the Universe to $\t = 6.52/h$ billion
years old. However for large, or even moderate $h$ estimates ($\gtrsim 0.65$)
these simplifying assumptions resulted in an age crisis in which the universe
was younger than our Galaxy ($\t \approx 10$ Gyr $< t_{Gal} \approx 12$ Gyr).

With a new inventory of the Universe described in the previous section and 
a new more precise value for the Hubble constant
(e.g. Mould \etal 2000), a new more precise age for the Universe can be calculated.
This was the focus of an article I recently published in Science entitled
 ``A Younger Age for the Universe''.
The result I obtained was more than a billion years younger than other recent
results (see Fig. 7).

\section{What could be wrong?}

Doubts about some of the observation used here are discussed in Dekel \etal (1998).
The contribution of neutrinos (or another form of hot dark matter) to $\om$  remains a wild card.
It is possible that supernovae are not as uniformly bright as we believe.
It is possible that the well-motivated assumptions used in the CMB analysis (gaussian adiabatic fluctuations, structure formation
through gravitational instability) are mistaken.
A mild conspiracy of unknown systematic errors could substantially change 
the constraints in Fig. 5F.

There has been some speculation recently that the evidence for $\ol$ is really 
evidence for some


\begin{figure}
\centerline{\psfig{figure=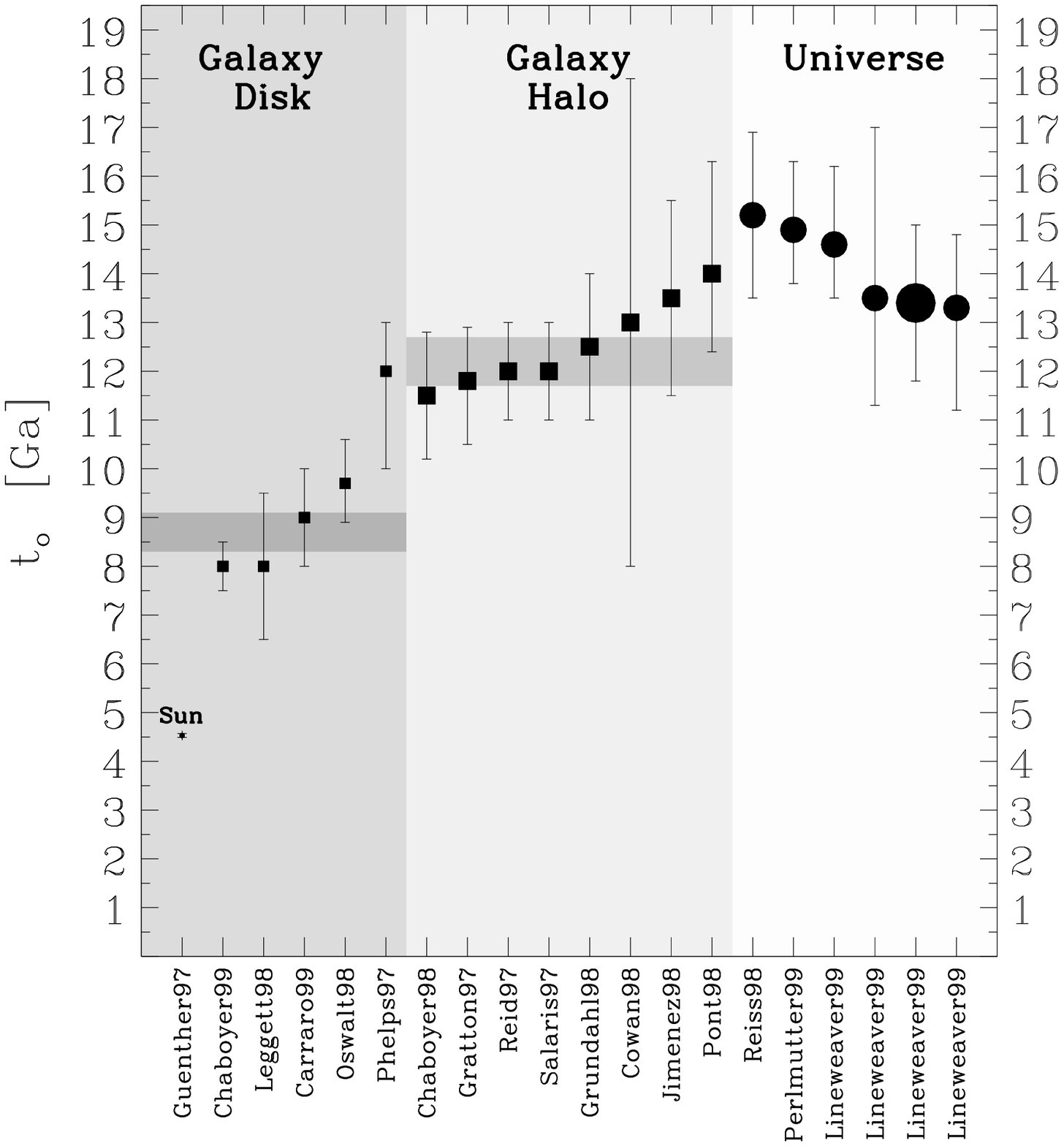,angle=0,height=13.5cm,width=14cm,rheight=14.5cm,rwidth=14cm}}
\vspace{-0.3cm}
\caption{ {\bf Figure 7. How old is the Universe?}\\
A compilation of recent age estimates for the Universe and for the oldest 
objects in our Galaxy.
Estimates of the age of the Universe are based on estimates of $\om$, $\ol$ and $h$. Galactic age estimates  
are direct in the sense that they do not depend on cosmology.
Averages of the estimates of the age of the Galactic halo and Galactic disk are shaded grey.
The absence of any single age estimate more than $\sim 2 \sigma$ from the average
adds plausibility to the possibly overdemocratic procedure of computing the 
variance-weighted averages.
The age of the Sun is accurately known and is included for reference.
The largest dot at $13.4 \pm 1.6$ Ga (billion years) is the main result of the Lineweaver (1999a) paper.
This age range is shaded grey on the x-axis of the Fig. 8.
Comfortingly, the Universe is older than the objects in it. 
This has not always been the case in cosmology and its absence has been a leading cause 
of cosmology bashing.
} \label{fig-7}
\end{figure}

\clearpage
\begin{figure}
\centerline{\psfig{figure=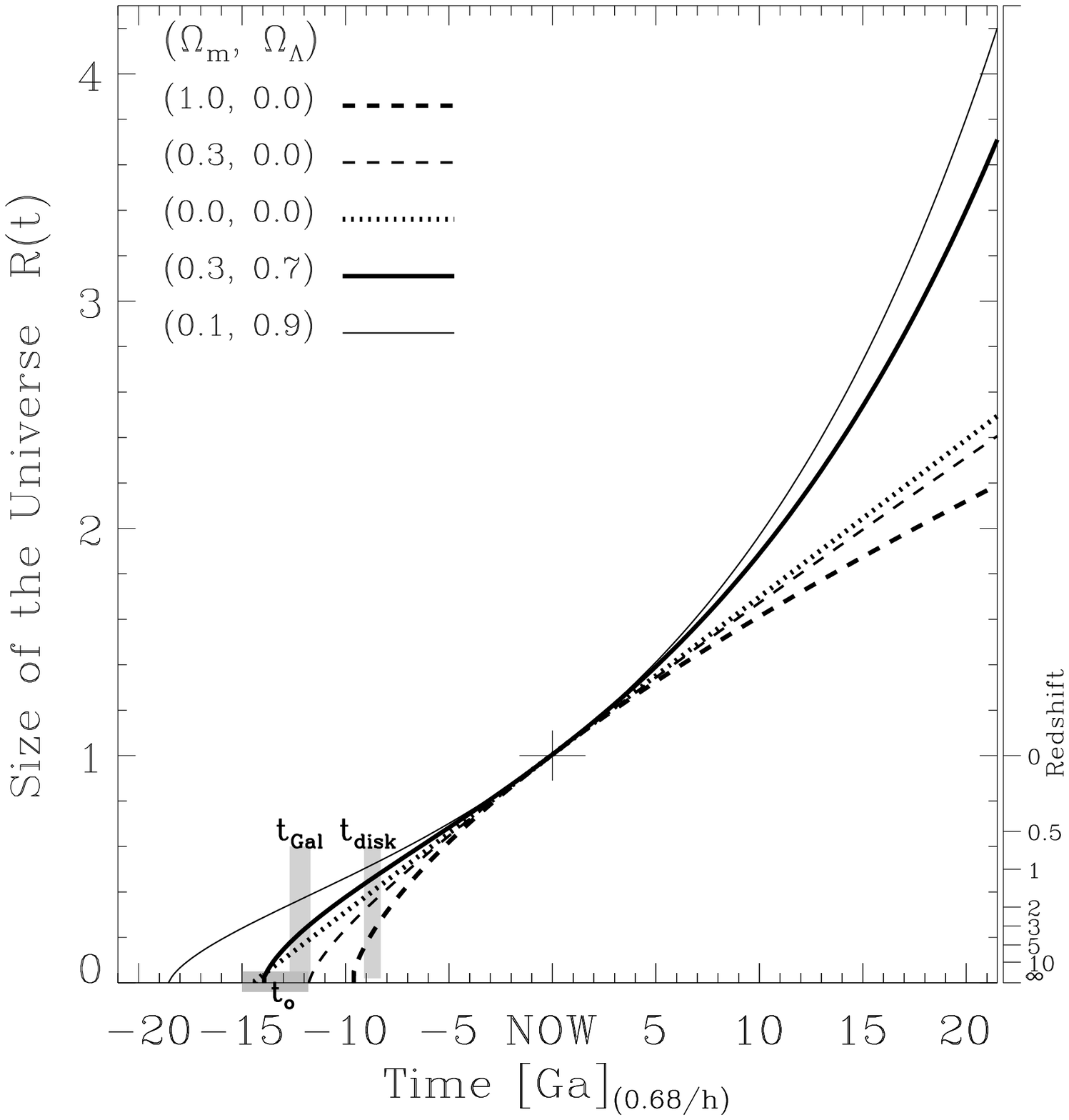,angle=0,height=14cm,width=14cm}}
\caption{{\bf Figure 8. Destiny of the Universe.}\\
The size of the Universe, in units of its current size, as a function of time.
The age of the five models can be read from the x axis as the time between `NOW'
and the intersection of the model with the x axis. The main age result from Lineweaver 1999a,
$\t = 13.4 \pm 1.6$ Gyr, is labeled $\t$ and is shaded grey on the x axis.
Models containing $\ol$ curve upwards ($\ddot{R} > 0$) and are currently accelerating.
The empty universe has $\ddot{R} = 0$ (dotted line) and is `coasting'. 
The expansion of matter dominated universes is slowing down ($\ddot{R} < 0$).
The $(\ol, \om) = (0.3, 0.7)$ model is favoured by the data.
Over the past few billion years and  on into the future, the rate of expansion
of this model increases. This acceleration means that we are in a period of slow 
inflation.
} \label{fig-8}
\end{figure}

\noindent
form of stranger dark energy (dubbed `quintessence') that we have been incorrectly 
interpreting as $\ol$. Several workers have tested this idea. The evidence so far 
indicates that the cosmological constant interpretation fits the data as well as 
or better than an explanation based on more mysterious dark energy 
(Perlmutter \etal 1999a, Garnavich \etal 1998, Perlmutter \etal 1999b).

\section{The Future}

As the quality and quantity of cosmological data improve, 
the questions: What is the Universe made of? How old is the Universe?
will get increasingly precise answers from an ever-tightening network of constraints.
An army of instruments is coming on line.
Better CMB detectors are being built; long duration balloons will fly; 
sensitive new high resolution interferometers will soon be on line and 
we all have high expectations for the
two CMB satellites  MAP and Planck.
In the near future, new CMB measurements will 
reduce the error bars in Fig. 5 by a factor of $\sim 5$
and/or... if some inconsistency is found, force us to change our basic 
understanding of the Universe. Maybe inflation is wrong, 
maybe CDM doesn't exist or we live in an eternally inflating  multiverse.

The biggest prize of all may be something unexpected.
We know that our model of the Universe is incomplete at the largest
scales and that it breaks down as we get closer and closer to the big bang.
It seems very probable that our model is wrong in some unexpectedly 
fundamental way. It may contain some crucial conceptual blunder 
(as has happened so many times in the past).
Some unexpected quirk in the data may point us in a new direction
and revolutionize our view of the Universe on the largest scales. 
Surely this is the golden age of cosmology.

What does this all mean for the physicists in the street?  
We should devote more effort to studying nothing -- the vacuum.
We should improve on measurements of the Casimir effect.
Maybe one of us will invent a heat engine based not on a phase transition
of water but on a phase transition of the vacuum.
In the past, on the few occasions where general relativity and 
quantum theory intersected ($\ol$ is a quantum term in a classical equation)
exciting new things have emerged:  Hawking radiation, 
entropy calcuations of black holes
and maybe soon a theoretical calculation of $\ol$ which will
lead to a plausible theory of quantum gravity.

\bsk
\baselineskip = 12pt
{\abstract \ni ACKNOWLEDGMENTS
I am supported at UNSW by a Vice Chancellor's Research Fellowship.
I acknowledge the editorial support of Kathleen Ragan, author of
``Fearless Girls: Heroines in Folktales from Around the World'' }

\clearpage
\baselineskip = 12pt
{\references \ni REFERENCES
\ssk

\ref Cen, R. \& Ostriker, J.P. 1999, \apj, 514, 1-6 
\ref Cohn, J. 1998, Astrophysics and Space Science, v. 259, 3, p. 213-234 (also available at http://xxx.lanl.gov/abs/astro-ph/9807128)
\ref Dekel, A., Burstein, D. \& White, S.D.M. 1998, in ``Critical Dialogues in 
Cosmology'' ed. N. Turok (World Scientific:River Edge, NJ) pp 175-191
\ref Fukuda, Y. \etal 1998, Phys. Rev. Lett. 81, 1562
\ref Fukugita, M. Hogan, C., Peebles, P.J.P. 1998, \apj, 503, 518
\ref Garnavitch, P.M. \etal 1998, \apj, 509, 74
\ref Harrison, E. ``Cosmology, the Science of the Universe'', 1981, Cambridge Univ. Press
\ref Lineweaver, C. H. \& Barbosa, D. 1998, \apj, 496, 624
\ref Lineweaver, C. H. 1998, \apj, 505, L69
\ref Lineweaver, C. H. 1999a, Science, 284, 1503, also available at http://nedwww.ipac.caltech.edu/level5/cos\_par.html
\ref Lineweaver, C. H. 1999b, in ``Gravitational Lensing: Recent Progress and Future Goals'', edt T. Brainerd \& 
C. Kochanek, available at http://xxx.lanl.gov/abs/astro-ph/9909301
\ref Mould J.R. \etal in preparation, available at http://xxx.lanl.gov/abs/astro-ph/9909260
\ref Perlmutter, S. \etal 1999a, \apj, 517, 565 
\ref Perlmutter, S., Turner, M.S., White, M. 1999b, available at http://xxx.lanl.gov/abs/astro-ph/9901052
\ref Riess, A.G. \etal 1998, A.J. 116, 1009, (also available at http://xxx.lanl.gov/abs/astro-ph/9805201)
\ref Smoot, G.S. \etal 1992, \apj, 396, L1 
\ref Turner, M.S. in press, available at http://xxx.lanl.gov/abs/astro-ph/9904051
}

\section{Bio}
PHOTO\\
Charles Lineweaver is a Vice Chancellor's Research Fellow at the University of New South Wales.
He studied undergraduate physics at Ludwig Maximillian Universit\"at, Germany and
at Kyoto University, Japan.
He obtained his PhD in Physics from the University of California at Berkeley. After a postdoctoral fellowship at 
Strasbourg, France he came to UNSW in 1997.
He also has an undergraduate degree in history
and a masters degree in English.
He has lived or travelled in 58 countries, speaks four languages
and has played semi-professional soccer.
He is co-convener with John Webb of a popular new UNSW course on bioastronomy: ``Are We Alone?''.

\end{document}